\documentclass[sigconf]{acmart}
\AtBeginDocument{%
  \providecommand\BibTeX{{%
    \normalfont B\kern-0.5em{\scshape i\kern-0.25em b}\kern-0.8em\TeX}}}

\copyrightyear{2024} 
\acmYear{2024} 
\setcopyright{rightsretained} 
\acmConference[CHI EA '24]{Extended Abstracts of the CHI Conference on Human Factors in Computing Systems}{May 11--16, 2024}{Honolulu, HI, USA}
\acmBooktitle{Extended Abstracts of the CHI Conference on Human Factors in Computing Systems (CHI EA '24), May 11--16, 2024, Honolulu, HI, USA}
\acmDOI{10.1145/3613905.3650923}
\acmISBN{979-8-4007-0331-7/24/05}

\usepackage{multirow}
\begin{document}

\title[No Risk, No Reward]{No Risk, No Reward: Towards An Automated Measure of Psychological Safety from Online Communication}

\author{Sharon Ferguson}
\email{sharon.ferguson@mail.utoronto.ca}
\orcid{0000-0002-2091-3435
}
\affiliation{%
  \institution{Mechanical and Industrial Engineering, University of Toronto}
  \city{Toronto}
  \state{ON}
  \country{Canada}
}

\author{Georgia Van de Zande}
\affiliation{%
  \institution{Mechanical Engineering, Massachusetts Institute of Technology}
  \city{Cambridge}
  \state{Massachusetts}
  \country{USA}}

\author{Alison Olechowski}
\affiliation{%
  \institution{Mechanical and Industrial Engineering, University of Toronto}
  \city{Toronto}
  \state{ON}
  \country{Canada}
}






\renewcommand{\shortauthors}{Ferguson et al.}

\begin{abstract}
  The data created from virtual communication platforms presents the opportunity to explore automated measures for monitoring team performance. In this work, we explore one important characteristic of successful teams --- Psychological Safety --- or the belief that a team is safe for interpersonal risk-taking.  To move towards an automated measure of this phenomenon, we derive virtual communication characteristics and message keywords related to elements of Psychological Safety from the literature. Using a mixed methods approach, we investigate whether these characteristics are present in the Slack messages from two design teams--- one high in Psychological Safety, and one low. We find that some usage characteristics, such as replies, reactions, and user mentions, might be promising metrics to indicate higher levels of Psychological Safety, while simple keyword searches may not be nuanced enough. We present the first step towards the automated detection of this important, yet complex, team characteristic. 
\end{abstract}




\received{25 January 2024}
\received[accepted]{1 March 2024}

\maketitle

\section{Introduction}
With the increasing popularity of remote and hybrid working styles, teams are communicating and collaborating via computer-mediated tools more than ever. These teams are often dispersed, which makes supervision more challenging, as the team's dynamics and predictors of success are harder for managers to monitor. One important requirement for successful teams is Psychological Safety (PS) \cite{edmondson2023psychological, frazier2017psychological, newman2017psychological}. The concept of PS was popularized in 1999 as a team-level construct defined as the ``shared belief that a space is safe for interpersonal risk-taking'' \cite{edmondson1999psychological}. Since its inception, PS has been shown to promote 
the ability to learn from mistakes \cite{edmondson1999psychological, edmondson2023psychological}, share knowledge \cite{edmondson2023psychological, frazier2017psychological}, and achieve better overall performance \cite{edmondson1999psychological, edmondson2014psychological, newman2017psychological}.
Despite the widespread focus on PS across numerous research fields, it is almost exclusively measured via the seven-item survey measure developed in 1999 \cite{edmondson1999psychological, newman2017psychological}. Since the survey only captures perceptions at a single point in time, team leaders looking to monitor their team's dynamics over time must redistribute the survey regularly. The survey can also fail to measure the team-level construct if each team member has different perceptions. 

Thus, we argue that an automated approach to measuring this important team construct can be an early warning signal for teams and their leaders when harmful dynamics emerge and require intervention. As teams rely increasingly on computer-mediated communication, we propose that the data created on these platforms is an ideal starting point for automating the measurement of PS. We consider text-based messaging platforms, termed Enterprise Communication Platforms (ECPs) \cite{ferguson2023we}, as the source of data in this work, as this data reflects the team's work, language, and relationships. Predicting team performance from collaboration data has proven successful in past Human-Computer Interaction (HCI) research \cite{cao2021my, jung2016coupling, wang2022group}; for example, \citet{cao2021my}'s prediction of team viability via chat messages. 

Despite the success of past work, PS is a complex team construct, with various ways in which it can be displayed. Thus, its automated identification is a challenging undertaking. In this paper, we work towards this ultimate goal by addressing the following research question: \textbf{What characteristics of ECP communication might relate to high or low levels of Psychological Safety?} In this exploratory study, we analyze the ECP (Slack) messages from two hybrid design teams: one comparatively high in PS, and one comparatively low. We quantitatively characterize their use of Slack, including patterns of replies and emoji reactions, and qualitatively analyze conversations that may signal PS.

We find that Slack use characteristics, such as the number of replies, number and type of emoji reactions, and user mentions may be a starting point to consider in the automated prediction of PS, as they differed across the two teams studied here, whereas other Slack behaviours did not. Further, we found that the counts of keywords and phrases, which represent key concepts within PS (voice, supportive, learning, and familiarity behaviours), do not differentiate the teams in our sample. Qualitative analysis revealed that this was due to false positives and nuanced differences in the way that these terms are used by the high team compared to the low team.


We contribute the first step towards holistically understanding how PS may be detected and automatically measured on online platforms,
extending past work \cite{ahuja2019digital} which has only used a single linguistic category as a proxy for the phenomenon. We provide an initial understanding of the communication characteristics which may differentiate comparatively high and low PS teams. We bring this exploratory study to the HCI community to spur discussion on next steps and other phenomena that could be measured using collaboration data.

\section{Background}

Psychological Safety has been identified as a characteristic required for teams to innovate \cite{edmondson2023psychological, frazier2017psychological, newman2017psychological}, learn from failure \cite{newman2017psychological, carmeli2009high}, share knowledge \cite{edmondson2014psychological, edmondson2023psychological, frazier2017psychological, newman2017psychological}, and be satisfied \cite{frazier2017psychological}.
PS is similar to trust, although it extends to describe ``a team climate characterized by interpersonal trust and mutual respect in which people are comfortable being themselves'' [p.~354]\cite{edmondson1999psychological}. While the construct is primarily measured via the seven-item scale proposed in \citet{edmondson1999psychological}, scholars have divided PS into four categories to observe it: voice (or silence), supportive (or unsupportive), learning, and familiarity behaviours \cite{o2020measuring}. While PS was initially used to study successful healthcare teams
\cite{edmondson1999psychological}, the concept captured the attention of managers when it was found to be a key differentiator of successful teams in Google's Project Aristotle \cite{duhigg2016google}. 

PS has also been studied in virtual teams \cite{lechner2022create, hao2022trust, kirkman2021building}, and is believed to be a key component for resilient virtual teams \cite{kirkman2021building}. For example, \citet{lechner2022create} found that virtual teams are less likely to build PS than their in-person counterparts are, so they must invest in cultivating it. \citet{hao2022trust} found that the level of virtuality in a team strengthened the relationship between PS, trust and knowledge-sharing. \citet{edmondson2023psychological} acknowledge the importance of studying how remote work will affect team PS, and they call for methodological innovations to do so. 

PS is one important characteristic of successful teams, and much past HCI work looks to predict team success overall. For example, \citet{jung2016coupling} predicts team performance from conflict instances, \citet{Borge2012patterns} from team interaction patterns, and \citet{zhou} from team feedback. Researchers have recently also turned to virtual collaboration platforms as a data source when predicting team success. \citet{wang2022group} focused on how successful teams organized Slack chats. \citet{cao2021my} used features of chat conversations to predict team viability, and they describe a set of their predictors as relating to PS, though they are not derived from the definition of the construct. 

We explore the possibility of predicting one component of team success, PS, using virtual communication data.

\section{Methods}

As this work is exploratory, we cast a wide net, collecting a broad set of virtual communication characteristics that may represent PS. We use both quantitative text analysis and qualitative thematic analysis to validate these characteristics. We detail common categories of PS from past work.
As the foundational survey in \cite{edmondson1999psychological} is the most widely accepted and validated measure of PS, we began by identifying the behavioural characteristics from these items. The survey refers to making mistakes, bringing up tough issues, taking risks, accepting others, asking for help, being supportive, and valuing skills. As this is the most common measure of PS \cite{newman2017psychological}, most other studies also build from this definition --- such as interview and observational studies \cite{o2020measuring, grailey2021exploring}. Ultimately, we found that the categories within the observation scheme by \citet{o2020measuring} act as a comprehensive summary of the key parts of the PS definition: \textbf{voice}, \textbf{supportive (or unsupportive)}, \textbf{learning}, and \textbf{familiarity behaviours}. We augmented the definition of each of these categories (defining the sub-categories ourselves) from other published measures of PS \cite{edmondson1999psychological, edmondson2014psychological, grailey2021exploring}. For example, we placed 
the elements of the definition of PS within each of these buckets: discussing mistakes and issues is part of \textbf{voice} and accepting, supporting and valuing others are part of \textbf{supportive}. Further, \citet{edmondson1999psychological} identifies how PS leads to \textbf{learning} behaviour --- such as asking for feedback, or providing suggestions to improve a piece of work. Lastly, teams which are high in PS often display their comfort with each other using familiarity behaviours \cite{o2020measuring, roberto2002lessons}, such as jokes \cite{potipiroon2021does}. Emojis have also been shown to only be used once teams have developed a level of familiarity \cite{shandilya2022need}. These elements, and their definitions, are summarized in Table \ref{tab:definitions}. 

\begin{table*}[]
\centering
\caption{Categories of Psychological Safety and associated keywords and Slack behaviours. Categories from \cite{o2020measuring}. .* represents regex match for any characters.}
\label{tab:definitions}
\resizebox{\textwidth}{!}{%
\begin{tabular}{p{3.5cm}p{2.5cm}p{6.5cm}p{5.5cm}}
\textbf{Category} &
  \textbf{Sub-Category} &
  \textbf{Keywords and Phrases in Communication Messages} &
  \textbf{Slack Behaviours} \\ \hline
\multirow{4}{*}{Voice Behaviours} &
  Mistakes &
  sorry, mistake, apolog.* &
  \multirow{4}{5.5cm}{less time to reply, less time between messages, file share (incl. URLs), many replies, edits} \\
 &
  Critiques &
  incorrect, disagree, wrong, impossible, unlikely, "don't think" &
   \\
 &
  Asking for Help &
  "don't know", unsure, help&
   \\
 &
  Questions &
  who, what, where, why, how, ? &
   \\ \hline
\multirow{2}{*}{Supportive Behaviours} &
  Agreement &
  yes, yeah, ya, yea, agree.* &
  \multirow{2}{5.5cm}{Many replies, many emoji reactions, equality of messages sent} \\
 &
  Appreciative &
  congrat.*, amazing, amaze, wonderful, wow, thank.* &
   \\ \hline
Unsupportive Behaviours &
  Unappreciative &
  "not needed", stop, waste &
  Few replies, few emoji reactions, non-equality of messages sent \\ \hline
\multirow{2}{*}{Learning Behaviours} &
  Suggestions &
  improv.*, better, instead, actual.*, "what if" &
  \multirow{2}{*}{@mentions, @channel} \\
 &
  Asking for Input &
  feedback, share, thoughts, idea.* &
   \\ \hline
Familiarity Behaviours &
  Emojis &
  ;{[}\textasciicircum{}/\textasciicircum{}\textbackslash{}\textbackslash{}\textasciicircum{}{]}*?: &
  emoji reactions \\
 &
  Jokes &
  hah.*, aha.*, lol.*, lmao.*, jok.* &
   \\ \hline
\end{tabular}
}
\end{table*}

We created a list of representative keywords and phrases, that may be found in Slack communication, for each sub-category (Table \ref{tab:definitions}). 
\textbf{Voice behaviours} represent team members feeling safe to share opinions, ideas, critiques, mistakes or misunderstandings with their team \cite{edmondson2014psychological, edmondson2023psychological, newman2017psychological}. 
We identified question words (``who'', ``what'', etc.), critique words (``incorrect'', ``disagree'', etc.), requests for help (``help'', ``unsure", etc.), and mistakes (``sorry'', ``mistake'', etc.) in this category. We consider Slack behaviours such as attachments or message edits (which could signal hesitancy to share), and the number of and time to reply (as reactions to voice behaviour are important). \textbf{Supportive behaviours} are also important reactions to \textbf{voice behaviours} \cite{newman2017psychological}. They are defined by agreement and appreciation \cite{o2020measuring}. We characterized these with agreement terms (``yes'', ``yeah'', etc.) and appreciation (``congrat.*'', ``wow'', etc.). \textbf{Supportive behaviours} may also be correlated with replies or emoji reactions to show support for ideas that someone shared --- thus we look at the number and type of emoji reactions, and the equality of contributions on Slack. We describe \textbf{unsupportive behaviours} as the opposite of \textbf{supportive behaviours}, characterized by unappreciative terms (``not needed'', ``stop'', etc.). \textbf{Learning behaviours}, such as looking to improve or asking for feedback, also characterize teams high in PS \cite{edmondson1999psychological}. We have defined these with suggestion terms (``improv.*'', ``better'', etc.) and requests for input (``ideas'', ``thoughts'', etc.). We also measure user or channel mentions, as these requests for feedback may be targeted. Lastly, teams high in PS exhibit \textbf{familiarity behaviour} \cite{o2020measuring}, characterized by emojis (in text or reactions), or jokes (``hah.*'', ``lol.*'').  

We used regular expression keyword searches for these words and phrases within each team's Slack messages. We tabulate the number of messages within each team that contain these keywords. We also qualitatively analyze the messages containing these keywords to further validate these choices and elaborate on the quantitative results. 
We used a qualitative thematic analysis \cite{braun2006using} process to identify similarities and differences in how these keywords were used between the teams. The first author applied open codes to each individual message in the high team and the low team. Then, the codes were consolidated (aggregated and organized hierarchically) in each of the four categories \textbf{voice, supportive/unsupportive, learning,} and \textbf{familiarity.} Then, these codes were compared between the high team and the low team, identifying which codes were common across both teams, and which represented differences. 

Further, we tabulate the counts and averages of the Slack behaviours between the two teams.

To validate these identified characteristics, we study the entire public channel (i.e., not direct messages) Slack communications from two design teams. This data contains the content of the messages as well as metadata, such as author, timestamp, attachments, replies, and emoji reactions (emoji ``stickers'' that can be used to react to a message). We study two of the six student teams that participated in a Mechanical Engineering capstone course at a major US Institution in 2023. The teams contained 14--16 students who collaborated with over 15 staff members to develop a functional physical product within three months. The course represents a realistic, though condensed, design process where teams manage their schedules and budgets, procure materials, and design and manufacture the product. This uniform context, where team members come from the same student population and are given the same resources, allows us to control for a number of external factors that may influence PS, such as organization culture \cite{edmondson2014psychological}, or leadership style \cite{frazier2017psychological, kahn1990psychological} so that the communication characteristics can be studied more clearly. Further, these teams are close to industrial teams 
as many products go on to be patented or commercialized. These teams are hybrid --- they meet and work together in person but also communicate virtually regularly.
Each team is designated a Slack workspace. More details about these teams can be found in \cite{ferguson2022communication, ferguson2023we}. 

We surveyed these teams thrice throughout the semester to capture their perceptions of PS \cite{edmondson1999psychological}.
The surveys were administered directly on Slack. The survey response rates were moderate, anywhere from 10\% to 65\% (per-team, per-period). While PS is a team-level construct, in cases where the team construct fails to be measured (either from incomplete data, or misaligned perceptions) researchers measure \textit{perceptions of Psychological Safety} instead \cite{letting2023obstacles, schulte2012coevolution} -- which is the average of the available PS questionnaire responses. We measure \textit{perceptions of Psychological Safety} but refer to it as Psychological Safety (PS) in the remainder of this paper. 

\begin{figure}[h]
\includegraphics[width=0.5\textwidth]{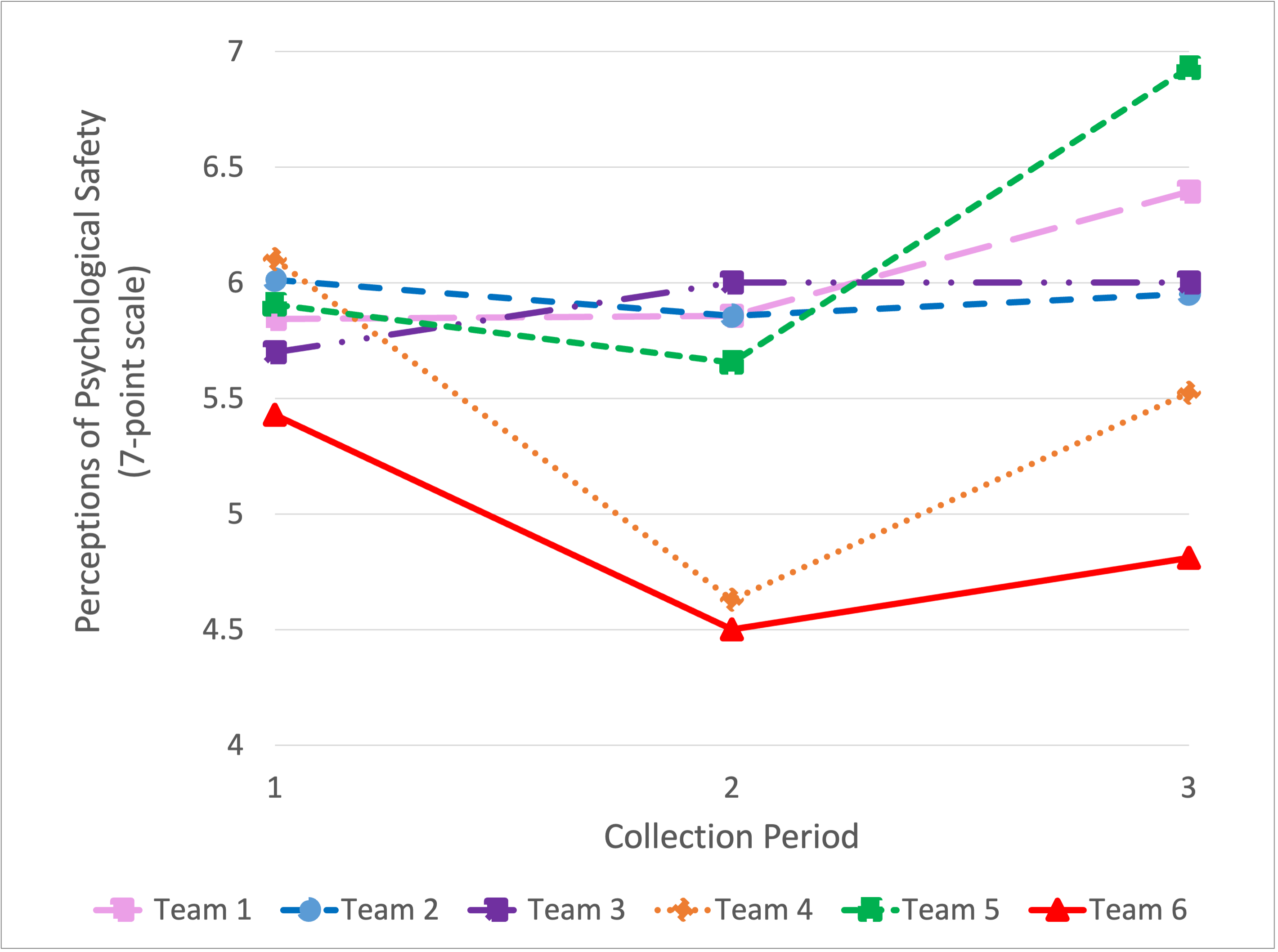}
\caption{Average perceptions of Psychological Safety values per each team over the semester.}
 \Description{A scatter plot with 6 different teams. The X-axis represents 3 data collection periods, and the Y-axis represents the 7-point Psychological Safety scale. Team 6 consistently shows the lowest levels of Psychological Safety (between 4.5-5.5) across all three periods of data collection, and Team 2 shows the most consistently high value, around 6. }
\label{fig1}
\end{figure}

We chose our two teams based on consistently high or consistently low PS scores.
Figure \ref{fig1} shows all teams' perceptions of PS across time periods. We identified that Team 6 (in red) consistently had the lowest PS. While Team 5 had the highest average across time periods, they experienced significant variation throughout the semester. Since we were studying a semester-wide process, 
we chose the team with a consistently high score, or the lowest standard deviation, Team 2 (in blue). Thus, we compare Team 2 and Team 6 in this exploratory paper. Individuals provided informed consent for the surveys and to share their Slack data. Anyone who did not consent to sharing their Slack data had all of their Slack interactions removed from the dataset (three individuals from Team 6 and one from Team 2).

\section{Results}

\subsection{Quantitative Tabulations}

We tabulated the Slack behaviours for the high and low PS teams, as seen in Table \ref{tab:my_label}. We comment only on the characteristics which differ between teams. The low PS team sent more public channel messages than the high team. Within \textbf{voice behaviours}, the high team sent more replies, and a higher proportion of their messages had at least one reply. 
Within \textbf{supportive behaviour}, the high team used more emoji reactions than the low team (also a factor of \textbf{familiarity behaviour}). Slack contribution was also more equal in the high team, with a standard deviation of 3.4\% of total messages, compared to 6.4\% in the low team.
In \textbf{learning behaviours}, the high team used @user mentions slightly more often than the low team. This suggests that number of replies may be a promising avenue in signalling \textbf{voice} and \textbf{supportive behaviour}, number of emoji reactions and equality of online communication may contribute to \textbf{supportive behaviour}, and user mentions may signal teams' \textbf{learning behaviour}. 

\begin{table*}
\centering
\caption{Count of Slack behaviours. Percentages represent the percentage of total messages sent.}
\label{tab:my_label}
\resizebox{\textwidth}{!}{%
    \begin{tabular}{p{5cm}p{2cm}p{2cm}p{5cm}p{2cm}p{2cm}}
        Slack Behaviour & High Value & Low Value & Slack Behaviour & High Value & Low Value\\ \hline
        
        Total Number of Messages & 5494  & 7496 & \textbf{Supportive Behaviour} & & \\
        \textbf{Voice Behaviours} & & & Number of Emoji Reactions & 4857  & 2279\\
        Number of Replies & 3897 & 3340 &
         \% of Messages With at Least One Emoji Reaction &  23\%& 18\% \\
         \% of Messages With at Least One Reply & 18.27\% & 12.91\% 
         & Standard Deviation of \% Contribution to Total Slack Messages & 0.034 & 0.064\\
          Average Time to Reply & 3:35:59& 3:43:10 
        & \textbf{Learning Behaviours} & &  \\
        
        Average Time Between Messages in a Channel & 18:33 & 17:18 & Number of @Channel Mentions & 41 (1\%)& 160 (2\%) \\
        Number of File Shares & 990 (18\%)& 1162 (16\%) & Number of @User Mentions & 539 (10\%) & 504 (7\%)\\
        Number of Edits & 243 (4\%) & 357 (5\%)  \\
        
        \bottomrule
    \end{tabular} 
    }
\end{table*}

As the high PS team used more emoji reactions than the low team did, we were interested in the type of emojis that were used. Table \ref{tab:emojis} lists the ten most commonly used emoji reactions per team. The teams studied are assigned colours, and we see team members use the heart emojis in their team's colour often, perhaps to build team spirit. A custom emoji of the photo of one team member was the second most commonly used emoji reaction on the high team, perhaps representing team bonding. 
Five of the top ten emoji reactions were custom in the high team, while only three were custom in the low team. This could represent more \textbf{familiarity behaviour} in the high PS team, such as inside jokes. 

\begin{table*}
    \centering
    
    \caption{Top ten most used emoji reactions for each team. $\dagger$ represents a custom-made emoji. Percentages are based on the total amount of emoji reactions used.}
    \resizebox{\textwidth}{!}{%
    \begin{tabular}{p{1cm}p{4cm}p{4cm}p{1cm}p{4cm}p{4cm}}
        Rank & High Team & Low Team& Rank & High Team & Low Team\\ \hline
        1 & Team colour heart (1657, 34\%) & Team colour heart (749, 33\%) & 6 & Team spinner$\dagger$ (146, 3\%)& Course emoji$\dagger$ (84, 4\%)\\
        2 & Team member photo$\dagger$ (422, 9\%) & Thumbs up (541, 24\%) &  7 & Heart (143, 3\%) & Laughing (83, 4\%)\\
        3 & Thumbs up (356, 7\%) & Heart (156, 7\%) & 8 & Team colour pet$\dagger$ (138, 3\%) & Eyes (58, 3\%)\\
        4 & White and green check mark (206, 4\%) & Cat ``party''$\dagger$ (126, 6\%) & 9 & Team colour fire$\dagger$ (126, 3\%) & Exclamation mark (56, 2\%)\\
        5 & Team ``party''$\dagger$ (185, 4\%) & Raised hands (105, 5\%) & 10 & Laughing (116, 2\%) & Team mascot$\dagger$ (33, 1\%)\\

        \bottomrule
    \end{tabular}
    \label{tab:emojis}
    }
\end{table*}

We identified the percentage of messages within each team that fell into each sub-category from Table \ref{tab:definitions}, shown in Table \ref{tab:tabulations}. We find that the high and low teams do not seem to differ in frequency in any of the sub-categories --- all values are within a few percent of each other. Keyword searches can hide nuance in how words are used \cite{schwartz2013personality}; thus, we conducted a follow-up qualitative analysis to identify if there are patterns between how these terms are used across teams.

\begin{table}[]
\centering
\caption{Number and percentage of total messages containing keywords within each sub-category of Psychological Safety.}
\footnotesize
\label{tab:tabulations}
\begin{tabular}{llll}
\textbf{Concept}                       & \textbf{Sub-concept} & \textbf{High Team} & \textbf{Low Team} \\ \hline
\multirow{4}{*}{\textbf{Voice Behaviours}}      & Mistakes             & 82 (1\%)               & 76 (1\%)               \\
                                       & Critiques            & 32 (<1\%)                & 51 (<1\%)               \\
                                       & Asking for Help      & 232  (4\%)              & 229 (3\%)              \\
                                       & Questions            & 1876  (34\%)             & 2339  (31\%)            \\ \hline

\multirow{2}{*}{\textbf{Supportive Behaviours}} & Agreement            & 241  (4\%)              & 391  (5\%)             \\
                                       & Appreciative         & 244   (4\%)             & 244  (3\%)             \\ \hline
\textbf{Unsupportive Behaviours}                & Unappreciative       & 22 (<1\%)                & 40 (<1\%)                \\ \hline
\multirow{2}{*}{\textbf{Learning Behaviours}}   & Suggestions          & 134  (2\%)              & 197 (3\%)               \\
                                       & Asking for Input     & 248  (5\%)              & 303 (4\%)              \\ \hline
\textbf{Familiarity Behaviours}                 & Emojis               & 295 (5\%)               & 317 (4\%)              \\
                                       & Jokes                & 85   (1\%)              & 101 (1\%)              \\ \hline
\end{tabular}
\end{table}

\subsection{Qualitative Observations}

\subsubsection{Voice Behaviours}
Within \textbf{voice behaviours}, we comment on mistakes, critique, asking for help, and questions. 

The word ``sorry'' was commonly used across teams when someone was late: \textit{``i might be about 30 min late..so sorry''}[high]. This may not be a high-stakes enough mistake to signal an interpersonal risk and correspondingly high PS. There was more variety in reasons why the individuals in the low team apologized, such as for asking for clarification, \textit{``We talked about so many things I got confused so sorry''} [low]; or for not communicating clearly, \textit{``sorry..I said that in the wrong channel oops''} [low].
Both teams also discussed mistakes regarding the product or design process, which might be a better signal of PS, as they have a larger impact on the team's success. For example, \textit{``I just came into lab and found that the soldering iron was left on...No need to claim this mistake..''}[high]; \textit{``I'd like to apologize because I assumed our [product] was running off 120VAC.''}[low]. In both these cases, a tool or product part was referenced --- perhaps future iterations of this algorithm need to filter for a combination of mistake keywords with task-specific terms to improve the detection of PS. Further, a wider variety of apology types (apologies combined with communication words or questions) may signal low PS, as it could represent a team that does not feel like they can ask clarifying questions without apologizing. 

We found elements of critiques of the team's process and product in Slack messages.
The low team more commonly critiqued the process, \textit{``..There is no path forward for [this product]; we don't have a sketch...have done zero testing. It will be almost impossible...''}[low]. On the other hand, product critiques were more common in the high team, \textit{``actually [I don't know] if we have enough pins to run a 3 digit 7-bit display...?''}[high]. While it undoubtedly can also be scary to critique the team's process, this could be perceived by team members more personally, eroding PS in the team over time. Thus, critique words paired with task-focused terms, might better signal high PS.

Within the category of asking for help, we caught instances when someone said they did not know something or needed help with it. We found that in the high team, this often happened at the end of a message, after the team member had shared some research or work they had done, or an idea they had. The author would share information, and then bring up a lack of knowledge as a roadblock, where they needed their teammates' help to move forward: `\textit{`@user <link> this site talks about dipoles over ground...I don't know nearly enough about antennas to try and solve the issue theoretically}''[high]. On the other hand, individuals on the low team tended to admit they didn't know something in response to a question asked or when feedback was sought: \textit{``I think that would be good for buoyancy but...I don't know how good it'll be at cushioning''}[low]. Thus, high PS might be signalled when the uncertainty or request follows information sharing, which may be characterized by attachments, URLs, or long descriptions. This could be the team members showing a ``good faith'' effort to try to solve the problem themselves first.

Questions were common among both teams, 
and we found no difference in how they were asked. 

\subsubsection{Supportive Behaviours}
Under \textbf{supportive behaviours}, we discuss agreement and appreciation. 
Since the project requires the team to make many decisions, both teams frequently referenced agreement. 
We found that mentions of agreement from the high team were mostly agreeing to a specific decision or proposition: \textit{``I agree that we should buy all 4..''}[high], and while the low team also had these instances of agreement, they commonly also questioned agreement: \textit{``I just wanted to be clear about this..are we all agreed that the ideal user is...''}[low]. Thus, agreement combined with uncertain language might signal low PS teams where the members may want to disagree, but feel they can't.

Both teams showed appreciation for their peers. It was more common in the high PS team to call out a specific person for a specific action, such as, \textit{``big thanks to [names] for helping clean up our space yesterday''}[high]. On the other hand, the low team shared more general compliments, \textit{``Slideshow looks amazing! Great job everyone!''}[low]. 

\subsubsection{Unsupportive Behaviours}
The keywords used for \textbf{unsupportive behaviours} were not effective at capturing instances of people underappreciating someone's effort. They resulted primarily in false positives such as saying someone ``stopped by'', discussing waste produced by the product, or stopping as a function the product should do. These phrases only captured a few instances where teams were reiterating perhaps unsupportive feedback they had received from outside the team. Thus, these keywords are not effective signals without further refinement.




\subsubsection{Learning Behaviours}
Under \textbf{learning behaviours} we discuss instances where teams shared suggestions or asked for input from others. In general, both teams shared suggestions about the product or the process, and we noticed no distinct patterns. There were also false positives, such as someone correcting themselves, telling a teammate to feel better, or asking about meeting logistics: \textit{``I'm currently in [location] but can move up if that's better?''} [high]. Perhaps combining these keywords with opinion marking terms such as ``I think'' could help to prevent these false positives. 

When teams asked each other for input, they often either requested feedback, thoughts, edits, votes, or ideas. 
The high team displayed more vulnerability in their asks, as they explicitly welcomed feedback or edits \textit{``..but def[initely] feel free to make edits''}[high], whereas the low team was more likely to make general calls for thoughts or ideas: \textit{``Any thoughts on a section name or mascot? Been saying [mascot name]...but more than happy to tweak if there's something you think would be better''}[low]. Perhaps directly asking for a teammate to critique or edit your work represents a stronger belief that the team is a safe space for risk-taking, and we should search for these vulnerability signalling terms.

\subsubsection{Familiarity Behaviour}
\textbf{Familiarity behaviour }is characterized by jokes and emojis. While there were many instances of jokes in both teams' communication, particularly making fun of themselves,
the product, 
or referencing inside jokes,
we found no difference in the way jokes were made between the high and low teams. 

Emojis were also used \textit{within} messages (note that these are the same set of emojis as emoji reactions, but a different use case). Teams commonly used their team colour emoji, and crying, smiling or laughing faces. Both teams used them in a self-deprecating way \textit{``you're telling me I left RIGHT when this started :sob:''}[high], and a supportive way \textit{``FANTASTIC job everyone with the posters and pitches!! :baby\_chick::hatching\_chick:''} [low]. However, emojis were mostly used in the low team to direct their team's attention towards important information: \textit{``:alert: URGENT UPDATES...:alert:''}[low]. This practice was less common in the high team, perhaps suggesting that the low team had poor communication practices (which can both result from \cite{newman2017psychological}, or cause \cite{siemsen2009influence}, low PS) and had to use other methods to capture attention.

\section{Discussion}

In this exploratory study, we aimed to identify characteristics of virtual communication that may differentiate between higher and lower levels of PS in teams, suggesting promising directions for future investigation.
Our quantitative analysis revealed that the number of replies, number and type of emoji reactions, and number of user mentions 
should be further investigated. We also found that simple keyword lists, alone, are not likely to be effective signals of PS. Qualitative analysis identified that this was due to nuance in how teams use these terms, such as the low team using emojis primarily to capture attention, whereas the high team used them in supportive messages.

Our results also highlight the complexity of measuring team-wide social phenomenon via text. 
Voice behaviour is an important sign of high PS \cite{o2020measuring}. While we can identify the presence of voice behaviour, text-based measures miss when someone chooses to remain silent, which signals lower levels of PS. Further, while all of the PS sub-categories are described in literature \cite{edmondson1999psychological, edmondson2014psychological, edmondson2023psychological, newman2017psychological, o2020measuring}, we did not detect differences in the amount of corresponding keyword-textual evidence in the teams studied here. First, the small 1-2\% differences we saw could increase if more teams were studied. Another possible interpretation is that there are different levels of PS --- as suggested by its scale measure \cite{edmondson1999psychological}--- and that these categories are enacted differently at different levels of PS. This is best illustrated by our finding that the high team asks for feedback and edits while the low team asks for thoughts and ideas. Even the low team rated themselves above the midpoint on the PS scale, which may explain why we see evidence of some learning behaviours, but not all. In future work, we will define not the presence or absence of a behaviour, but the level to which it is displayed. 

This exploratory study relied on only two teams (over 40 individuals and 13,000 messages). Only the messages sent by consenting team members were included in the analysis; thus, conversations reviewed manually may be missing messages. We also measure perceptions of PS based on the survey responses from a portion of the team, meaning that we may not have a complete picture of PS within the team. Despite these limitations, this exploratory work identified promising directions for the next iteration of an automated approach to detecting PS from virtual communications.

To further advance the automation of this team-wide measurement, we plan to test the communication patterns identified on a larger set of teams. We will continue iterating upon these keywords lists using the insights derived in this work (such as combining agreement terms with uncertainty measures). We will manually disentangle messages displaying PS from false positives, refine our search, and create a large, labelled dataset to validate our algorithm. We will experiment with more sophisticated detection methods, such as language models and network measures. 

As we continue working towards an automated measure of PS, there are a number of validity concerns to keep in mind. While here we begin by searching for text characteristics corresponding with the foundational definition of PS, future work is needed to ensure construct validity --- that the behaviours captured in the algorithm actually represent PS. To move towards this, we will broaden the text characteristics studied to look for characteristics that correlate with survey measures of PS across a larger number of teams. We will further validate this using interviews with the teams studied, verifying the intent behind, or outcome of, instances of their Slack communication. Studying a broader number of teams across contexts, and verifying our proposed measure with team members will also help to address the concern of endogeneity. We will continue to calibrate the automated measure with the well-accepted survey measure of PS \cite{edmondson1999psychological}, and triangulate our new measures with interviews or observations. Lastly, we will engage social psychologists with deep expertise in PS in comparing these different measures of PS, and validating the proposed algorithm. 

As teams continue to rely on computer-mediated communication, HCI has focused on understanding and improving this experience \cite{10.1145/3411764.3445243, 10.1145/3491101.3514488}, and recently, using the data from these platforms to do so \cite{cao2021my, wang2022group}. We contribute to this effort by moving towards the automated identification of PS, a concept important for the success of virtual teams \cite{kirkman2021building}, yet rarely discussed in HCI (see \cite{park2021armed} for one example). By studying characteristics related to each category in this complex phenomenon, we provide a starting point for researchers to move towards automatically measuring successful team dynamics in a number of online environments, such as open-source software teams \cite{10.1145/2468356.2468382} or Reddit communities \cite{10.1145/3544549.3585671}. 

The successful development of an automated measure for PS would allow system designers to build dashboards into virtual communication platforms for teams to monitor and adjust their dynamics. For example, we imagine this tool to be most useful for managers and team leaders --- PS is a strong predictor of a number of team outcomes \cite{newman2017psychological, frazier2017psychological, edmondson2023psychological, edmondson2014psychological}; thus, leaders having the opportunity to monitor the team's levels of PS and identify drops would allow timely interventions to get the team back on track. However, any tool which provides monitoring capabilities comes with risks. Employees may feel like they are being surveilled and like they need to filter their communication, acting in opposition to PS. Team members must be taught to treat these values as potential signals of a problem which should be investigated further, not as absolute truth, and should be careful not to make critical assumptions based on these values. To better understand how the tool will be used by teams, as well as to test the validity of the algorithm, we plan to conduct extensive user tests with industry teams. 

\section{Conclusion}
Psychological Safety is a prerequisite for many positive team outcomes, yet its measurement is currently limited to perceptions measured at a single point in time. This exploratory study aims to move towards an automated measure of Psychological Safety for teams to monitor, based on data from the virtual collaboration platforms already used by most teams. Using a mixed-method study of the Slack messages from two design teams, we find that the number of replies, number and type of reactions, and the number of user mentions are promising characteristics to further study, while simple keyword searches through messages hide many nuances. We contribute a set of communication characteristics that are a promising scope of focus for future iterations of an automated Psychological Safety measure.

\end{document}